\pdfoutput=1
\documentclass[11pt]{article}

\usepackage[final]{acl}

\usepackage[textsize=footnotesize]{todonotes} 

\usepackage{times}
\usepackage{latexsym}

\usepackage[T1]{fontenc}

\usepackage[utf8]{inputenc}

\usepackage{lscape}
\usepackage{svg}
\usepackage{microtype}
\usepackage{subcaption}
\usepackage{inconsolata}

\usepackage{graphicx}
\usepackage{booktabs}
\usepackage{multicol}

\newcommand\boldgreen[1]{\textcolor{olive}{#1}}
\renewcommand\boldgreen[1]{{#1}}

\newcommand{\shrink}{\vspace{-0.5\baselineskip}}
\newcommand{\sshrink}{\vspace{-0.25\baselineskip}}

\newcommand{\mypar}[1]{\smallskip\pagebreak[3]\noindent\textbf{#1}~}

\title{On Correlating Factors for Domain Adaptation Performance}

\author{G\"{o}ksenin Y\"{u}ksel \\
  University of Amsterdam \\ Amsterdam, The Netherlands\\
  \texttt{Goksenin.Yuksel@student.uva.nl} \\\And
  Jaap Kamps \\
  University of Amsterdam \\ Amsterdam, The Netherlands \\
  \texttt{kamps@uva.nl} \\}

\begin{document}
\maketitle
\begin{abstract}
Dense retrievers have demonstrated significant potential for neural information retrieval; however, they lack robustness to domain shifts, limiting their efficacy in zero-shot settings across diverse domains. In this paper, we set out to analyze the possible factors that lead to successful domain adaptation of dense retrievers. We include domain similarity proxies between generated queries to test and source domains. Furthermore, we conduct a case study comparing two powerful domain adaptation techniques. We find that generated query type distribution is an important factor, and generating queries that share a similar domain to the test documents improves the performance of domain adaptation methods. This study further emphasizes the importance of domain-tailored generated queries. 
\end{abstract}



\shrink
\section{Introduction}
\sshrink

Dense retrieval (DR) methods are extremely fast and easy to index \citep{dense-survey}. However, these models have to solve the much more challenging task of mapping inputs independently to a meaningful vector space \citep{thakur-etal-2021-augmented}. Compared to other neural IR methods, they show poor zero-shot performance across novel domains \citep{beir}.

Hence, there is a need to better understand the conditions and factors that contribute to the (lack of) zero-shot performance. \citet{ren2023a} analyze various attributes affecting the zero-shot performance of the DR methods. They provide an in-depth empirical analysis of different factors on zero-shot retrieval performance

In this paper, we set out to analyze the variables that may contribute to a successful domain adaptation. We examine the possible correlating factors for successful domain adaptation from the standpoint of \citep{ren2023a}. We choose to carry out our analysis on the GPL \citep{wang2021gpl} method, as this method was the most successful domain adaptation method for dense retrievers without taking hybrid models into account \citep{ren2023a}. We further conduct a case study using the InPars \citep{inpars} method and test the applicability of our findings in cross-encoder domain adaptation.

We extend the factors of \citep{ren2023a} with 
various attributes regarding the generated queries, such as: 
generated query vocabulary overlap between source/test queries, 
generated query vocabulary overlap between source/test documents, 
generated query type distribution entropy, and  
generated query type distribution overlap between source/test queries.

Our main contributions are the following. 
First, we find that the GPL method transfers to LoTTE datasets and shows substantial improvements in NDCG@10 compared to the baseline. This indicates that the GPL method is a robust domain adaptation method that shows performance improvements over a broad spectrum of domains. 
Second, we discover that "Generated Query Type Entropy is one of the critical attributes that correlates positively with the performance of the domain-adapted retriever. 
Third, we further find that the overlap of the generated domain to the test domain is an indicator of domain adaptation performance regardless of the framework.


The rest of this paper continues with related work (\S\ref{sec:rel}), our approach and experimental setup (\S\ref{sec:met}), the corresponding results and analysis (\S\ref{sec:res}), and ends with discussion and conclusions (\S\ref{sec:dis}).

\shrink
\section{Related Research}
\label{sec:rel}
\sshrink

This section discusses earlier research on performance factors 
and on domain adaptation.

\mypar{Factors for zero shot performance}
To the best of our knowledge, \cite{ren2023a} is the first paper to analyze attributes affecting the zero-shot performance of dense retriever methods. They provide an in-depth empirical analysis by investigating the effect of different factors on zero-shot retrieval performance, such as query/passage vocabulary overlap, query-type distribution, query scale, and lexical bias.

\mypar{Domain Adaptation}
Query generation (QG) models generate synthetic queries from documents. They introduce synthetic training data by generating new queries to form a positive pair with the document from which they have been generated. This procedure introduces domain-specific synthetic data to fine-tune the model. QGen \cite{QGen} uses a transformer to create synthetic queries for the training data with binary relevance labels. GPL \cite{wang2021gpl} extends the work of QGen by using the cross-encoder model to generate pseudo labels for synthesized query-document pairs.

GPL utilizes QG and successfully adapts the dense-retriever to the domain by showing improvement on 18 of 19 BEIR datasets \cite{wang2021gpl}. Moreover, InPars~\cite{inpars} uses a large language model (LLM) to generate synthetic queries. They use OpenAI's public API to prompt GPT-3 to generate plausible queries for a document, showcasing that fine-tuning cross-encoders in this synthetic dataset yield state-of-the-art results in many datasets. Their research further reiterates that the generated synthetic queries provide a solid learning signal to fine-tune neural IR models. 

\shrink
\section{Method}
\label{sec:met}
\sshrink
This section details our experimental setup.

\mypar{Vocabulary Overlap} We consider each vocabulary's top 10K most frequent words (excluding stopwords). We calculate the percentage of vocabulary overlap by the weighted Jaccard \cite{ioffe} for each pair of query and document sets. 

\mypar{Query Type Distribution} Following \cite{ren2023a}, we first categorize each query and generate the query to a specific type depending on the keywords. The used keywords are in Appendix A.

Later, we measure the entropy of the query type distribution. Higher entropy yields a more uniform distribution and, thus, more diverse query types. Furthermore, we use cross-entropy to analyze the overlap of query-type distributions.

\mypar{Additional Factors}
We analyze the generated queries to understand their similarities to test and source domains. For this, we utilize generated query-source query, generated query-test query, generated query-source document, and generated query type distribution overlap between source/test queries.

We analyze the Spearman correlation for both models' improvement and retention capabilities with respect to the factors.\footnote{For the CQADupstack dataset, we take the average for all the factors}

\mypar{Models}
For BEIR tasks, we use the open source models and generated queries from \cite{wang2021gpl}. Moreover, we use the pipeline described in \cite{wang2021gpl} to adapt the GPL/msmarco-distilbert-margin-mse model on the LoTTE dataset. Furthermore, we use the generated queries provided by \citep{inpars}.


\mypar{BEIR} \citep{beir}
combines several existing datasets into a heterogeneous suite for “zero-shot IR” tasks spanning bio-medical, financial, and scientific domains. Our paper utilizes open-access datasets from BEIR, which has 14 out of 19 collections. 

\mypar{LoTTE} \citep{santhanam-etal-2022-colbertv2}
is a new dataset for Long-Tail Topic-stratified Evaluation for IR. To complement the out-of-domain tests of BEIR, LoTTE focuses on natural user queries that pertain to long-tail topics, ones that an entity-centric knowledge base like Wikipedia might not cover. LoTTE consists of 12 test sets, each with 500–2,000 queries and 100k–2M passages .

\begin{figure*}[!t]
    \centering
    \begin{subfigure}{\linewidth}
        \centering
        \includegraphics[width=0.95\linewidth]{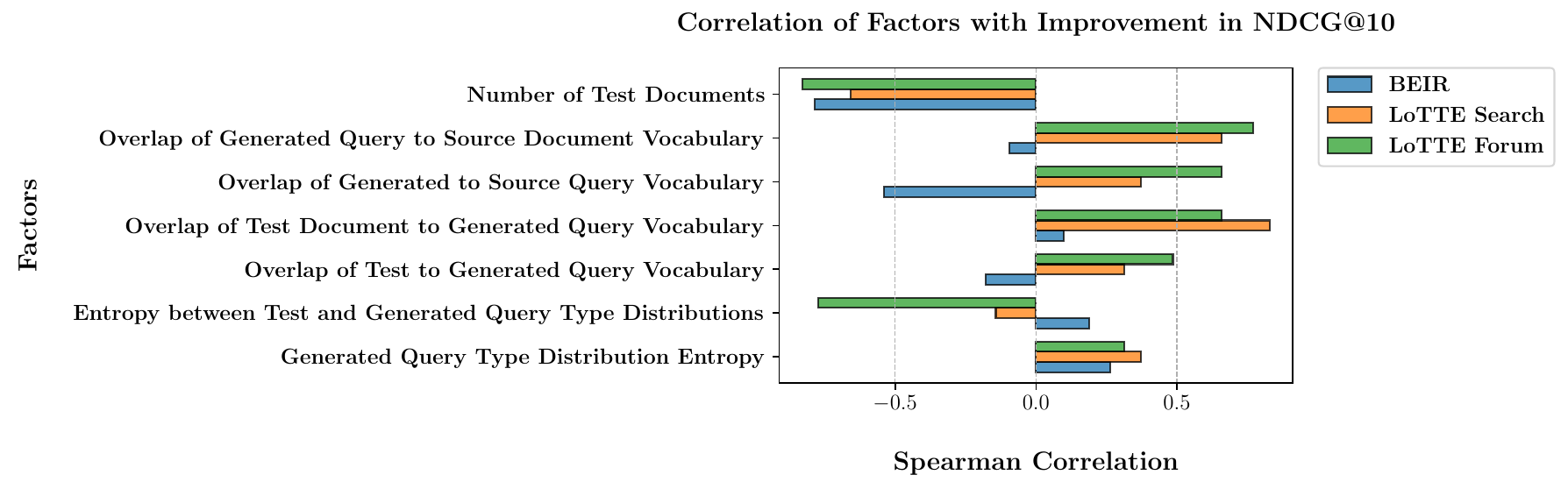}
        \caption{Factors affecting domain adaptation performance: GPL on different collections.}
        \label{fig:domain_adaptation_gpl}
    \end{subfigure}
    \begin{subfigure}{\linewidth}
        \centering
        \includegraphics[width=0.95\linewidth]{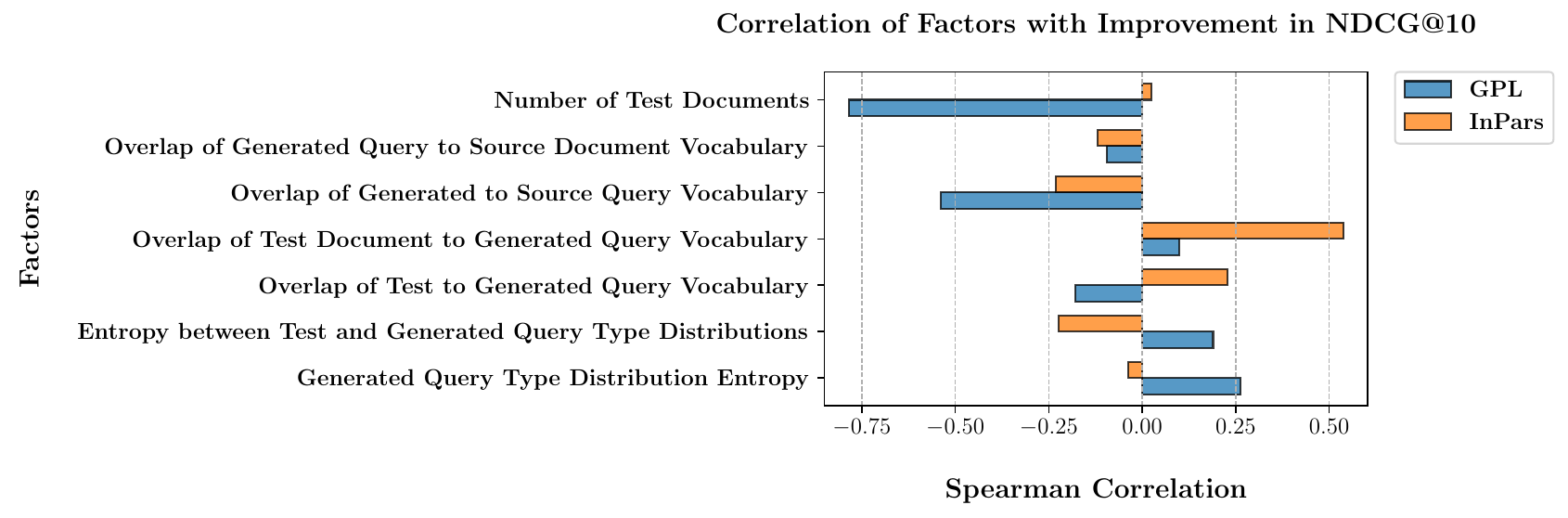}
        \caption{Factors affecting domain adaptation performance: GPL's T5 vs InPars query generation.}
        \label{fig:domain_adaptation_inpars}
    \end{subfigure}
    
    \caption{Comparison of factors affecting domain adaptation performance measured in NDCG@10 improvement.}
    \label{fig:domain_adaptation}
\end{figure*}

\shrink
\section{Results}
\label{sec:res}
\sshrink

This section discusses our results on the factors for successful domain adaptation.

\begin{table}[t]
\setlength\tabcolsep{0pt} 
\small
\begin{tabular*}{\columnwidth}{@{\extracolsep{\fill}} lccc}
\toprule
\bf Corpus & \textbf{BM25} & \textbf{MSMARCO} & \textbf{GPL}\\
\midrule
\multicolumn{4}{l}{\sl LoTTE Search Test Queries} \\

            \fontsize{8}{10}\selectfont\textbf{Writing} & 63.2 & 70.6 & $\textbf{77.1}^{\boldgreen{\fontsize{4}{6}+6.5}}$   \\
            \fontsize{8}{10}\selectfont\textbf{Recreation} & 59.8 & 62.8 & $\textbf{71.0}^{\boldgreen{\fontsize{4}{6}+8.2}}$  \\
            \fontsize{8}{10}\selectfont\textbf{Science} & 38.6 & 46.4 & $\textbf{52.0}^{\boldgreen{\fontsize{4}{6}+5.2}}$ \\
            \fontsize{8}{10}\selectfont\textbf{Technology} & 44.5 & 57.6 &$\textbf{64.1}^{\boldgreen{\fontsize{4}{6}+6.5}}$  \\
            \fontsize{8}{10}\selectfont\textbf{Lifestyle} & 68.1 & 77.0 & $\textbf{82.8}^{\boldgreen{\fontsize{4}{6}+5.8}}$\\
            \fontsize{8}{10}\selectfont\textbf{Pooled} & 52.4 & 62.1 & $\textbf{65.2}^{\boldgreen{\fontsize{4}{6}+3.1}}$ \\
            \midrule
            \multicolumn{4}{l}{\sl LoTTE Forum Test Queries}\\
            \fontsize{8}{10}\selectfont\textbf{Writing} & 66.5 & 66.8 & $\textbf{72.2}^{\boldgreen{\fontsize{4}{6}+5.4}}$  \\
            \fontsize{8}{10}\selectfont\textbf{Recreation} & 56.3 & 59.9 & $\textbf{66.8}^{\boldgreen{\fontsize{4}{6}+6.9}}$\\
            \fontsize{8}{10}\selectfont\textbf{Science} & 35.1 & 34.3 & $\textbf{39.7}^{\boldgreen{\fontsize{4}{6}+5.4}}$ \\
            \fontsize{8}{10}\selectfont\textbf{Technology} & 40.4 & 41.5 & $\textbf{50.0}^{\boldgreen{\fontsize{4}{6}+8.5}}$ \\
            \fontsize{8}{10}\selectfont\textbf{Lifestyle} & 62.4 & 69.3 & $\textbf{74.4}^{\boldgreen{\fontsize{4}{6}+5.1}}$\\
            \fontsize{8}{10}\selectfont\textbf{Pooled} & 48.3 & 52.4 & $\textbf{54.3}^{\boldgreen{\fontsize{4}{6}+1.9}}$ \\
\bottomrule
\end{tabular*}
\caption{Performance on LoTTE in NDCG@10.}
\label{tab:lotte}
\end{table}


\sshrink
\subsection{GPL Domain Adaptation}
\sshrink

GPL is indeed effective for domain adaptation.   We have reproduced the experiments of GPL on BEIR \citep{wang2021gpl}.  The results for the 14 public BEIR datasets are an NDCG@10 of 41.1 for BM25 and 41.8 for the zero-shot base model, and 46.4 (+4.6 point) for the domain-adapted GPT model.  

We extend these experiments to LoTTE in Table~\ref{tab:lotte} and observed consistent improvement over the base model.  The results for the 12 LoTTE datasets are an NDCG@10 of 53.0 for BM25 and 58.4 for the zero-shot base model, and 64.5 (+6.1 point) for the domain-adapted GPT model.

\sshrink
\subsection{Performance Factors}
\sshrink

Figure~\ref{fig:domain_adaptation_gpl} shows that \emph{generated query type distribution entropy correlates positively with the improvement in NDCG@10}. Models trained with more diverse generated query types show better domain adaptation performance in the target domain. 

Moreover, \emph{the entropy between test and generated query type distribution correlates negatively with NDCG@10 improvement}. The models trained with similar generated query types to target query types get better at retrieving documents in the target domain. 

We further find that \emph{vocabulary overlap of generated queries to test documents correlates positively with domain adaptation performance}. This suggests that generating queries with similar vocabulary to test documents is important for domain adaptation performance. As the generated queries share a similar domain to the testing domain, we observe better performance gains.

Moreover, \emph{the number of test documents correlates negatively with the improvement in NDCG@10}. This could be due to document subsampling. In GPL, the number of generated queries averages around 250k regardless of the corpus size. For smaller corpora, we generate more queries per document to match this requirement, resulting in full utilization of test documents. However, in larger datasets, we sub-sample, resulting in less coverage of target documents.

\begin{figure}[!t]
    \centering
    \begin{subfigure}{\linewidth}
        \includegraphics[width=\linewidth]{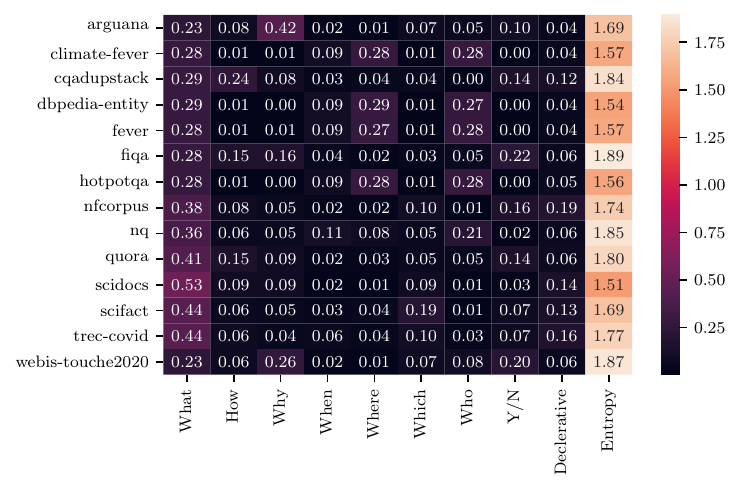}
        \caption{GPL}
        \label{fig:beir_queries}
    \end{subfigure}
    \hfill
    \begin{subfigure}{1\linewidth}
        \centering
        \includegraphics[width=1\linewidth]{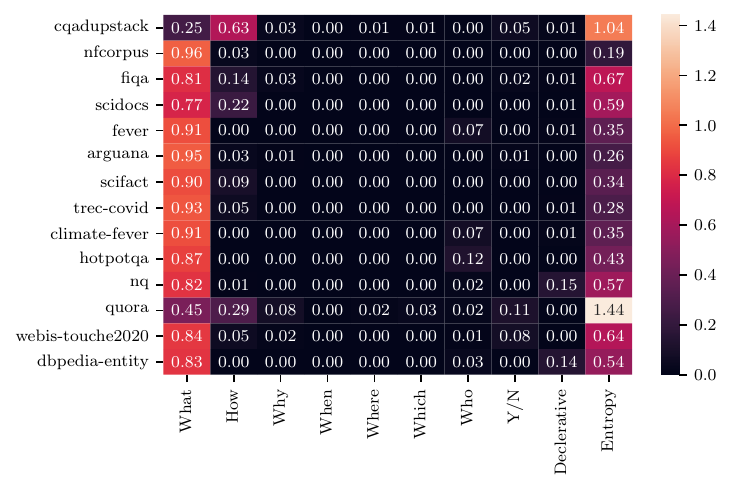}
        \caption{InPars}
        \label{fig:inpars_queries}
    \end{subfigure}
    \caption{Generated query type distribution for BEIR dataset}
    \label{fig:overall_comparison}
\end{figure}

\sshrink
\subsection{Case Study}
\sshrink
Figure~\ref{fig:domain_adaptation_inpars} shows the correlating factors for the domain adaptation performance in BEIR datasets from InPars, and GPL framework. For the InPars framework, the overlap of generated queries to the test domain is the only positive factor. Unlike the previous findings, the number of test documents and generated query type distribution entropy are not correlating factors for InPars. Moreover, the overlap of source domain to generated queries harms both frameworks.

Figures~\ref{fig:beir_queries} and~\ref{fig:inpars_queries} show the generated query type distributions from GPL and InPars.  InPars queries are less diverse than those from GPL and are biased towards "What" queries. For InPars, queries generated from Quora and CQADupStack sets have higher entropies.

\begin{figure}[!t]
    \centering
    \begin{subfigure}{\linewidth}
        \centering
        \includegraphics[width=1\linewidth]{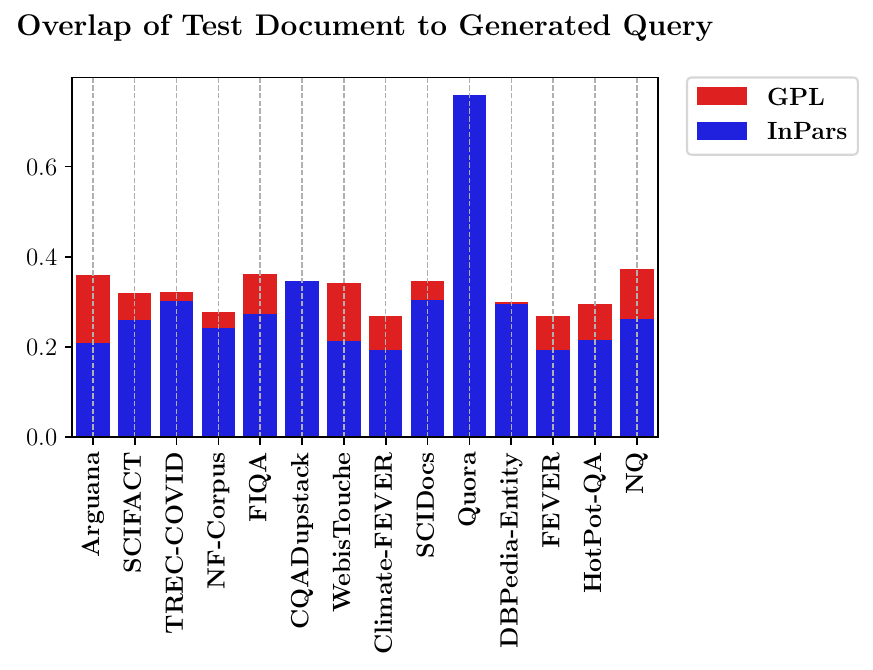}
    \end{subfigure}
    \begin{subfigure}{\linewidth}
        \centering
        \includegraphics[width=1\linewidth]{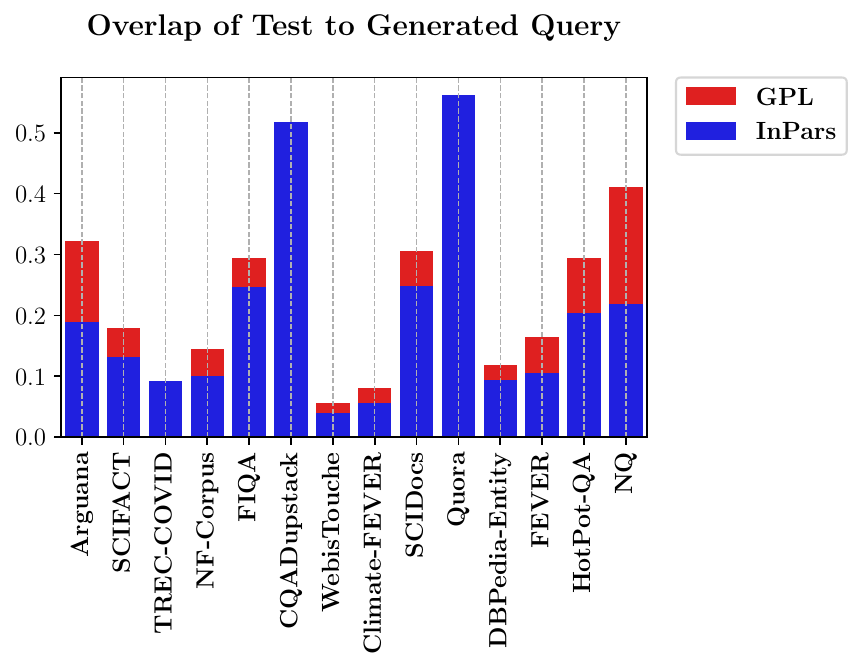}
    \end{subfigure}
    \caption{Factors for test to generated domain overlap in BEIR datasets. Two generated domain are compared,  GPL and InPars. }
\label{fig:overlap}
\end{figure}

Figure~\ref{fig:overlap} shows the attributes for measuring generated query to test domain overlap. InPars queries show substantial overlap with Quora and CQADupStack domains. Furthermore, GPL-generated queries overlap better with the test domains overall, especially visible for the NQ and Arguana domains.

\shrink
\section{Discussion and Conclusions}
\label{sec:dis}
\sshrink

This paper analyzed the contributing factors to domain adaptation performance for dense retrievers using BEIR and LoTTE datasets and conducted a case study comparing two state-of-the-art domain adaptation methods. Our findings show that generated queries are essential to domain adaptation performance, and generating diverse queries resembling the test documents likely improves performance for dense retrievers.

Additionally, our case study reveals that queries generated by the GPL framework have higher diversity. We find that generated queries from GPL overlap more with the test domain than those from InPars. Figures~\ref{fig:overall_comparison} and~\ref{fig:overlap} reveal these phenomena.

These findings could be attributed to differences in the training procedure of query generator models. The GPL framework uses MSMARCO trained T5 model to generate synthetic queries, and \citep{ren2023a} shows that the MSMARCO dataset has high query entropy. Whereas LLMs are not specifically trained to generate queries from a passage. 

Regardless of the framework, overlap between the test domain and generated queries is an indicator of domain adaptation performance. In contrast, similarity to the source domain harms the domain adaptation performance. This finding emphasizes the importance of well-generated synthetic datasets tailored to the domain regardless of the adapted model.
In future research, we plan on using the InPars framework to adapt dense retrievers and further analyze the contributing factors for such methodology.

\pagebreak[3]\shrink
\section{Limitations}
\sshrink

Although we investigate a diverse set of 26 corpora and tasks from BEIR and LoTTE, this is still restricted to English and to relatively common tasks.  It is of general importance investigate whether the similar factors play similar roles across languages and over different document genres and tasks.  In particular in the context of domain adaptation this is important, and this provides a means to understand the impact of language and domain shifts, and offers actionable ways to improve NLP performance under such diverse sets of conditions. 

Considering more diverse domains will increase the linguistic and task differences between the source and target domains, which may make the domain transfer for current domain adaption approaches far more challenging.  This may make the current domain adaptation approaches less effective in a setting where domain adaptation is most needed.  One attractive strategy would be to perform domain adaptation in several steps, by finding intermediate tasks and domains so we can transfer through a chain of models to bridge between the source and target domains. One could even consider combining multiple chains of models as in the original knowlegde distillation approach \citep{hinton2015distilling}. 

This paper looked at the impact of the generated queries in domain adaptation.  It is important to also consider the role of the pseudo-label oracle in the performance of domain adaptation.  In particular, it would be fruitful to see the importance of the source dataset of teacher model on the domain adaptation performance. GPL model uses knowledge distillation from the teacher model trained on MSMARCO datasets, which may be a limitation when considering an even more diverse sets of applications.  For future work, a comparison between different source datasets creating different teacher models would be helpful.  


\bibliography{main}

\appendix

\section{Query Types}

For queries starting with “WH” words, taking
“what” as an example, queries with the first word
of “what” or “what’s” are classified as what-type
queries. Queries starting with the first word “is”,
“was”, “are”, “were”, “do”, “does”, “did”, “have”,
“has”, “had”, “should”, “can”, “would”, “could”,
“am”, “shall”. are classified as Y/N queries. The
rest of the queries belong to declarative queries.

\newcommand{\anon}[1]{#1}
\renewcommand{\anon}[1]{\textsl{withheld to preserve anonymity}}

\section{Data, code, and trained models}
We share all our data, code, and pretrained models on GitHub (\anon{\url{XXX}}) and HuggingFace (\anon{\url{XXX}}). 

\end{document}